# Relativistic scattering with spatially-dependent effective mass in the Dirac equation

A. D. Alhaidari[a,1], H. Bahlouli[b], A. Al-Hasan[c], and M. S. Abdelmonem[b]

[a] *Shura Council, Riyadh 11212, Saudi Arabia*
[b] *Physics Department, King Fahd University of Petroleum & Minerals, Dhahran 31261, Saudi Arabia*
[c] *Samba Financial Group, Riyadh 11421, Saudi Arabia.*

We formulate an algebraic relativistic method of scattering for systems with spatially dependent mass based on the J-matrix method. The reference Hamiltonian is the three-dimensional Dirac Hamiltonian but with a mass that is position-dependent and having a constant asymptotic limit. Additionally, this effective mass distribution is locally represented in a finite dimensional function subspace. The spinor couples to spherically symmetric vector and pseudo scalar potentials that are short-range such that they are accurately represented by their matrix elements in the same finite dimensional subspace. We calculate the relativistic phase shift as a function of energy for a given configuration and study the effect of spatial variation of the mass on the energy resonance structure.

## I. INTRODUCTION

Calculations of physical quantities relevant to semiconductors are sometimes done using the effective mass approximation. Initially this approximation has been used to describe impurities in crystals [1] where much of the interaction with the host lattice is being parameterized through an effective mass parameter in the impurity Hamiltonian. The effective mass is also an important parameter in Landau's Fermi liquid theory that deals with low-level excited states of strongly interacting systems in a very appealing single particle approximation [2]. Since then, the effective mass approximation has become an essential ingredient in describing the transport properties of semi-conductor hetero-junctions and quantum dots [3]. One of the main features of these hetero-junctions or graded semiconductors is that the effective mass of the charge carriers is position dependent and is, frequently, the result of discontinuities across the hetero-junction with abrupt interfaces. Thus, one is lead to study quantum mechanical problems with position-dependent effective mass. However, such treatment encounters a nontrivial problem related to ordering ambiguity in the quantization of the momentum and mass operators in the kinetic energy term of the effective Hamiltonian.

On the other hand, relativistic effects have significant influence on the electronic properties of materials containing heavy atoms or those with heavy ion doping. This is because the charge carriers in such materials attain higher velocities comparable to lighter ones. Relativistic effects also include spin-orbit and spin-spin couplings which are purely relativistic corrections to the nonrelativistic Hamiltonians. Spin-orbit interaction, in particular, alters the spectroscopic properties of molecules containing heavy elements to a considerable extent. Thus the solution of the Dirac equation under the circumstance where the mass depends on the position of electrons will be of interest in studying materials containing heavy elements.

---
[1] E-mail: haidari@mailaps.org (corresponding author)

The quantization of non-relativistic Hamiltonians of position-dependent mass systems is always hindered by ordering ambiguities in the kinetic energy term. On the other hand its relativistic counterpart, the relativistic Dirac equation, does not suffer from such ambiguity. This ordering ambiguity of mass and momentum is due to the fact that these two quantities no longer commute when the mass is space dependent. An effective approach towards the resolution of this ambiguity is to start with the relativistic Dirac wave equation, which does not suffer from any ordering problem, then take the non-relativistic limit (which is well defined and unique) up to order $1/c^2$. There have been several attempts in defining the correct hermitian kinetic energy operator for a variable mass system based on current conservation [3], Galilean invariance [4] or the recent supersymmetric treatment of the effective mass Hamiltonians [5]. We believe that the work of Cavalcante et al. [6] is a measurable contribution towards the resolution of the ordering ambiguity problem of the quantum kinetic-energy operator with spatially varying effective mass. In our present work we opted for an algebraic method of quantum scattering, the J-matrix method, to deal with the scattering problem for space dependent mass systems. This method will enable us to obtain a highly accurate numerical solution of the relativistic scattering problem with space dependent mass distribution (while avoiding the ordering ambiguity) that is confined to a finite region in space but asymptotically constant.

The J-matrix is an algebraic method of quantum scattering developed almost thirty years ago [7]. The method exploits the fact that the unperturbed reference problem can be solved analytically in a certain complete set of $L^2$ basis functions. The basis set is chosen such that the matrix representation of the unperturbed Hamiltonian is tridiagonal (Jacobian). This property enables us to employ the analytical power associated with orthogonal polynomials. However the tridiagonal requirement of the reference Hamiltonian restricts the type of $L^2$ basis functions only to those that allow for such a tridiagonal representation of the unperturbed Hamiltonian [8]. Under these circumstances, the eigenvalue equation of the reference Hamiltonian gives rise to three-term recursion relation for the expansion coefficient of the unperturbed wave function. The short range scattering potential is then confined to an *N*-box in function space so that its matrix elements are zero outside this box (i.e., $V_{nm} = 0$ for $n, m \geq N$, where *N* is some large enough integer). This approximation provides us with a numerical mean to extract the necessary scattering information relevant to the problem at hand. Thus, the J-matrix structure in function space parallels that of the R-matrix method in configuration space [9]. A relativistic extension of the J-matrix has been proposed by Horodecki [10] for the Coulomb free interaction and by Alhaidari et al. [11]. It is the purpose of the present paper to formulate an algebraic relativistic method of potential scattering for systems with position dependent mass based on the J-matrix method.

The rest of our paper is organized as follows: in Sec. II, we give some preliminaries about the 3D Dirac Hamiltonian with position-dependent mass and define our reference Hamiltonian. In Sec. III, the J-matrix solution of the reference problem in the asymptotic region will be presented. The scattering matrix and associated phase shift are obtained in Sec. IV. Discussion of the numerical results obtained for a single channel scattering model is presented in section V.

## II. PRELIMINARIES

We consider the Dirac Hamiltonian in 3+1 dimensions with spherical symmetry and position-dependent mass $M(r)$ where the spinor couples to vector and pseudo-scalar potentials. That is, we study the three-dimensional relativistic problem with the following radial Dirac Hamiltonian

$$H = \begin{pmatrix} m(r) + \lambdabar^2 V(r) & i\lambdabar\left[\frac{\kappa}{r} + W(r) - \frac{d}{dr}\right] \\ -i\lambdabar\left[\frac{\kappa}{r} + W(r) + \frac{d}{dr}\right] & -m(r) + \lambdabar^2 V(r) \end{pmatrix}, \qquad (1)$$

where $\kappa$ is the spin-orbit quantum number with values $\pm 1, \pm 2, ...$, and it is related to the orbital angular momentum quantum number $\ell$ by $\kappa = \pm\left(\ell + \frac{1}{2}\right) - \frac{1}{2}$. $\lambdabar$ is the Compton wavelength $\lambdabar = \hbar/m_0 c = 1/c$. We have used the atomic units $\hbar = m_0 = 1$ where $m(r)$ is a dimensionless positive function that is related to the position-dependent mass by $m(r) = M(r)/m_0$. $V(r)$ and $W(r)$ are the radial "vector" and pseudo-scalar potential functions, respectively. Our choice of units over the conventional relativistic units (where $\hbar = c = 1$) is made to allow us to take the nonrelativistic limit, $c \to \infty$ ($\lambdabar \to 0$), in a very simple, intuitive, and straight-forward manner which is not possible in the latter units since $c = 1$. Additionally, it is easier to compare our results with those in atomic physics since the same system of units are used. Note also that $mc^2 \to \infty$ is not a good measure of the nonrelativistic limit for position-dependent mass systems because this limit could be satisfied in regions where $m \to \infty$ despite that the system might be highly relativistic (e.g. for systems with singular mass distribution, such as $m \sim \frac{1}{r}$). The time-independent wave equation to be solved is $(H - \varepsilon)\Psi = 0$, where $\varepsilon$ is the relativistic energy and $\Psi$ is the two-component spinor, which we write for positive energy as $\begin{pmatrix} i\psi^+ \\ \psi^- \end{pmatrix} e^{-i\varepsilon t}$. We study the class of problems in which we can write $m(r) = 1 + \lambdabar^2 S(r)$ [12]. In these cases, $S(r)$ plays the same role as that of a scalar potential in problems with constant mass. The Dirac equation could, thus, be written as

$$\begin{pmatrix} 1 + \lambdabar^2 V_+ - \varepsilon & \lambdabar\left(\frac{\kappa}{r} + W - \frac{d}{dr}\right) \\ \lambdabar\left(\frac{\kappa}{r} + W + \frac{d}{dr}\right) & -1 + \lambdabar^2 V_- - \varepsilon \end{pmatrix} \begin{pmatrix} \psi^+ \\ \psi^- \end{pmatrix} = 0, \qquad (2)$$

where $V_\pm = V \pm S$.

We can study several types of scattering problems depending on the choice of potentials to be included in the reference Hamiltonian such that the reference problem is exactly solvable and the matrix representation of the reference Hamiltonian is tridiagonal. The remaining potentials, $W$ and/or $V_\pm$, that are not included in the reference Hamiltonian are assumed to be short-range. That is, $W(r)|_{r>R} = 0$ and/or $V_\pm(r)|_{r>R} = 0$, respectively, where $R$ is the effective range of the potentials. For example, we could take the following reference Hamiltonian

$$\begin{pmatrix} 1 & \lambdabar\left(\frac{\kappa}{r}+W-\frac{d}{dr}\right) \\ \lambdabar\left(\frac{\kappa}{r}+W+\frac{d}{dr}\right) & -1 \end{pmatrix}, \tag{3}$$

where $V_{\pm}(r)|_{r>R}=0$. Nonetheless, in our present work we choose the free Dirac Hamiltonian as reference. That is, we take

$$H_0 = \begin{pmatrix} 1 & \lambdabar\left(\frac{\kappa}{r}-\frac{d}{dr}\right) \\ \lambdabar\left(\frac{\kappa}{r}+\frac{d}{dr}\right) & -1 \end{pmatrix}, \tag{4}$$

where $V_{\pm}(r)|_{r>R}=0$ and $W(r)|_{r>R}=0$. Therefore, the reference Hamiltonian results in the free Dirac wave equation, $(H_0-\varepsilon)\Phi(r,\varepsilon)=0$. Writing $\Phi(r,\varepsilon)$ in terms of its two spinor components as $\Phi=\begin{pmatrix}\phi^+\\\phi^-\end{pmatrix}$, this equation becomes

$$\begin{pmatrix} 1-\varepsilon & \lambdabar\left(\frac{\kappa}{r}-\frac{d}{dr}\right) \\ \lambdabar\left(\frac{\kappa}{r}+\frac{d}{dr}\right) & -1-\varepsilon \end{pmatrix}\begin{pmatrix}\phi^+\\\phi^-\end{pmatrix}=0, \tag{5}$$

which gives in the "kinetic balance" relations [13]

$$\phi^{\mp} = \frac{\lambdabar}{\varepsilon\pm 1}\left(\frac{\kappa}{r}\pm\frac{d}{dr}\right)\phi^{\pm} \tag{6}$$

resulting in the uncoupled Schrödinger-like second order differential equations

$$\left[\frac{d^2}{dr^2}-\frac{\kappa(\kappa\pm 1)}{r^2}+\frac{\varepsilon^2-1}{\lambdabar^2}\right]\phi^{\pm}(r,\varepsilon)=0. \tag{7}$$

There are two independent scattering solutions (where, $|\varepsilon|>1$) of this equation, one of them is regular (at the origin) and the other is not. For the upper spinor component, these solutions are written in terms of the spherical Bessel and Neumann functions as follows [14]:

$$\phi^+_{reg}(r,\varepsilon) = \sqrt{\frac{2}{\pi}}(\mu r)\begin{cases} j_\kappa(\mu r) &, \kappa>0 \\ j_{-\kappa-1}(\mu r) &, \kappa<0 \end{cases} \tag{8a}$$

$$\phi^+_{irr}(r,\varepsilon) = \sqrt{\frac{2}{\pi}}(\mu r)\begin{cases} n_\kappa(\mu r) &, \kappa>0 \\ n_{-\kappa-1}(\mu r) &, \kappa<0 \end{cases} \tag{8b}$$

where $\mu^2=(\varepsilon^2-1)/\lambdabar^2$. Near the origin, these upper components behave as follows: $\phi^+_{reg}\to\begin{cases}r^{\kappa+1},\kappa>0\\r^{-\kappa},\kappa<0\end{cases}$ and $\phi^+_{irr}\to\begin{cases}r^{-\kappa},\kappa>0\\r^{\kappa+1},\kappa<0\end{cases}$. On the other hand, asymptotically ($r\to\infty$) they are sinusoidal, $\phi^+_{reg}\to\sqrt{\frac{2}{\pi}}\times\begin{cases}\sin(\mu r-\pi\kappa/2),\kappa>0\\\cos(\mu r+\pi\kappa/2),\kappa<0\end{cases}$ and $\phi^+_{irr}\to\sqrt{\frac{2}{\pi}}\times\begin{cases}-\cos(\mu r-\pi\kappa/2),\kappa>0\\\sin(\mu r+\pi\kappa/2),\kappa<0\end{cases}$. To obtain the lower spinor component, $\phi^-$, we use the "kinetic balance" relation [13] and the differential properties of the Bessel and Neumann functions [14] giving

$$\phi^-_{reg}(r,\varepsilon) = \sqrt{\frac{2}{\pi}}\sqrt{\frac{\varepsilon-1}{\varepsilon+1}}(\mu r)\begin{cases} j_{\kappa-1}(\mu r) &, \kappa>0 \\ -j_{-\kappa}(\mu r) &, \kappa<0 \end{cases} \tag{9a}$$

$$\phi^-_{irr}(r,\varepsilon) = \sqrt{\frac{2}{\pi}}\sqrt{\frac{\varepsilon-1}{\varepsilon+1}}(\mu r)\begin{cases} n_{\kappa-1}(\mu r) &, \kappa>0 \\ -n_{-\kappa}(\mu r) &, \kappa<0 \end{cases} \tag{9b}$$

Note that in the nonrelativistic limit ($\hbar \to 0$, $\varepsilon \to 1 + \hbar^2 E$, where $E$ is the nonrelativistic energy) the lower components vanish. It should also be noted that the choice $\varepsilon = -1$ is not allowed in the kinetic balance relation (6) used to obtain $\phi^-$. Now, since $\varepsilon = -1$ belongs to the negative energy spectrum, then this implies that the solution obtained above does not include the negative energy solution of the relativistic problem. To obtain this negative energy solution one has to solve Eq. (7) for $\phi^-$ and use the *dual* kinetic balance relation $\phi^+ = \frac{\hbar}{\varepsilon - 1}\left(\frac{\kappa}{r} - \frac{d}{dr}\right)\phi^-$ to obtain $\phi^+$. It is, however, easy to verify that one solution is obtained from the other by the map $\varepsilon \to -\varepsilon$, $\kappa \to -\kappa$, and $\phi^+ \leftrightarrow \phi^-$. Therefore, in this work and from this point forward we will only be considering the positive energy solution of the problem. Using the orthogonality property of the Bessel functions one may verify that the regular solution is energy normalized in the sense that $\langle \Phi_{reg} | \Phi'_{reg} \rangle = \langle \phi^+_{reg} | \phi'^+_{reg} \rangle + \langle \phi^-_{reg} | \phi'^-_{reg} \rangle = \frac{2\varepsilon}{\varepsilon+1}\delta(\mu - \mu') = 2\hbar\sqrt{\frac{\varepsilon-1}{\varepsilon+1}}\delta(\varepsilon - \varepsilon')$. However, the irregular solution is not square integrable (with respect to the integration measure, $dr$). In the following section we utilize the J-matrix method to obtain the asymptotic solution of the reference problem.

## III. TRIDIAGONAL J-MATRIX REPRESENTATION: KINEMATICS

Due to the higher degree of symmetry of the reference problem, it is sometimes possible to find a special square integrable two-component spinor basis, $\{\chi_n\}_{n=0}^{\infty}$, such that the matrix representation of the reference wave operator, $H_0 - \varepsilon$, is tridiagonal for all $\varepsilon$. Precisely, $\langle \chi_n | H_0 - \varepsilon | \chi_m \rangle = J_{n,m}(\varepsilon)$ such that $J_{n,m} = 0$ for $|n - m| > 1$, where $n,m = 0,1,2,...$ This will allow for an algebraic solution of the reference wave equation for continuous values of the energy; a property which is desirable for scattering. The diagonal representation, on the other hand, can only admit discrete eigenenergies that are compatible with bound states. However, because the basis is square integrable and regular everywhere, then a faithful representation in this basis could only be obtained for the regular solution of the reference problem. As for the irregular solution, we can never hope to match its behavior everywhere using this basis. Nonetheless, it is still possible that this could be done asymptotically where it matters most for the scattering problem. Therefore, we construct two independent functions as infinite series in terms of the spinor basis $\{\chi_n\}_{n=0}^{\infty}$. We write one of them as $\Phi_{\sin}(r, \varepsilon) = \sum_n s_n(\varepsilon)\chi_n(r)$ and call it the "sine-like" solution. A "cosine-like" solution, on the other hand, is written as $\Phi_{\cos}(r, \varepsilon) = \sum_n c_n(\varepsilon)\chi_n(r)$, where the expansion coefficients $\{s_n\}_{n=0}^{\infty}$ and $\{c_n\}_{n=0}^{\infty}$ are independent. The sine-like solution will be identified with the regular solution of the reference $H_0$–problem, $\Phi_{reg}$. Therefore, the "sine-like" expansion coefficients $\{s_n\}_{n=0}^{\infty}$ will be obtained in terms of orthogonal polynomials that satisfy the three-term recursion relation resulting from the matrix-equivalent reference wave equation, $\sum_m J_{nm} s_m = \sum_{m=n,n\pm1} J_{nm} s_m = 0$. Subsequently, we show that these sine-like expansion coefficients, $\{s_n\}_{n=0}^{\infty}$, satisfy a second order linear differential equation in the energy. Hence, we find another independent set of solutions to this equation. These are precisely the "cosine-like" expansion coefficients $\{c_n(\varepsilon)\}_{n=0}^{\infty}$. However, we will find out that these expansion coefficients satisfy the same three-term recursion relation as $\{s_n\}$ except for the initial relation ($n = 0$). That is, $\sum_m J_{nm} c_m = 0$ for all $n \neq 0$. Precisely, $\sum_m J_{nm} c_m = \eta \delta_{n0}$, where $\eta$ is real and energy dependent. Therefore, the corresponding spinor wave function $\Phi_{\cos}(r,\varepsilon) = \sum_n c_n(\varepsilon) \chi_n(r)$, referred to also as the "regularized" wavefunction, does not satisfy the reference wave equation. It satisfies a regularized non-homogeneous wave equation that reads

$$(H_0 - \varepsilon)|\Phi_{\cos}\rangle = \eta(\varepsilon)|\tilde{\chi}_0\rangle, \tag{10}$$

where $\tilde{\chi}_0$ is an element of the set $\{\tilde{\chi}_n\}_{n=0}^{\infty}$ which is orthogonal to $\{\chi_n\}_{n=0}^{\infty}$ (i.e., $\langle \chi_n | \tilde{\chi}_m \rangle = \langle \tilde{\chi}_n | \chi_m \rangle = \delta_{nm}$). We emphasize again that since $\chi_n$ is $L^2$ and regular everywhere so is $\Phi_{\cos}(r,\varepsilon)$. However, asymptotically it is sinusoidal and identical to the irregular solution of the reference problem $\Phi_{irr}(r,\varepsilon)$. Therefore, the singular solution becomes regularized in the sense that it solves the modified wave equation (10).

Let us write $\chi_n = \begin{pmatrix} \varphi_n^+ \\ \varphi_n^- \end{pmatrix}$, then the following expression for $\varphi_n^+$ could be taken as a general $L^2$ function compatible with a series expansion of the upper component of the regular solution (8a)

$$\varphi_n^+(r) = e^{-x/2} \begin{cases} a_n^+ (\omega r)^{\kappa+1} L_n^{\nu_+}(x) & ,\kappa > 0 \\ a_n^- (\omega r)^{-\kappa} L_n^{\nu_-}(x) & ,\kappa < 0 \end{cases} \tag{11}$$

where $x = (\omega r)^\delta$, $L_n^\nu(x)$ are the Laguerre polynomials [14], and normalization constants are $a_n^\pm = \sqrt{\omega \Gamma(n+1)/\Gamma(n+\nu_\pm+1)}$. The basis parameters $\{\omega, \delta, \nu_\pm\}$ are real and such that $\omega > 0$, $\delta > 0$, $\nu_\pm > -1$. Here we choose to work in, what is referred as, the "Laguerre basis" where $\delta = 1$. However, in the Appendix we give solutions in the "oscillator basis" where $\delta = 2$. Now, since the spinor components of the regular solution (8a) and (9a) are related by the kinetic balance relation, then so too are the components of the basis. Therefore, the lower component of the spinor basis should be related to the upper as $\varphi_n^- = \frac{\hbar \omega}{\varepsilon + 1} \left( \frac{\kappa}{x} + \frac{d}{dx} \right) \varphi_n^+$. Using the differential relation and recursion properties of the Laguerre polynomials, $L_n^\nu(x)$, we obtain the following expression for the lower component of the spinor basis

$$\varphi_n^-(r) = \frac{1}{2} \frac{\hbar \omega}{\varepsilon + 1} e^{-x/2} \begin{cases} a_n^+ x^\kappa \left[ (4\kappa - \nu_+ + 1) L_n^{\nu_+} - (n+\nu_+) L_{n-1}^{\nu_+} + (n+1) L_{n+1}^{\nu_+} \right] & ,\kappa > 0 \\ a_n^- x^{-\kappa-1} \left[ -(\nu_- + 1) L_n^{\nu_-} - (n+\nu_-) L_{n-1}^{\nu_-} + (n+1) L_{n+1}^{\nu_-} \right] & ,\kappa < 0 \end{cases}$$

$$= \frac{1}{2} \frac{\hbar \omega}{\varepsilon + 1} x^{|\kappa|} e^{-x/2} \begin{cases} a_n^+ \left[ (n+\nu_+) L_n^{\nu_+-1} + (n+1) L_{n+1}^{\nu_+-1} + 2(2\kappa + 1 - \nu_+) L_n^{\nu_+} \right] & ,\kappa > 0 \\ -a_n^- \left( L_n^{\nu_-+1} + L_{n-1}^{\nu_-+1} \right) & ,\kappa < 0 \end{cases} \tag{12}$$

This expression shows that the basis in this representation is energy-dependent; a property which is not desirable from numerical point of view. This is because any calculation in such a basis has to be repeated for all energies in the range of interest. Nonetheless, we will find out shortly that this basis is only necessary for the analytic solution of the reference problem that ends at finding the coefficients $\{s_n, c_n\}$. Numerical

–6–

computations, on the other hand, will be carried out in a different, finite, and energy independent basis that will be given in the following section. Therefore, calculation in that energy independent basis (e.g., diagonalization of the Hamiltonian, phase shift analysis, etc.) will be done once and for all energies.

Now, the matrix representation of the free Dirac operator $H_0 - \varepsilon$ in the energy-dependent basis, $\{\chi_n\}$, becomes

$$J_{n,m} = \langle \chi_n | H_0 - \varepsilon | \chi_m \rangle = (1-\varepsilon)\langle \varphi_n^+ | \varphi_m^+ \rangle + (1+\varepsilon)\langle \varphi_n^- | \varphi_m^- \rangle \quad , \tag{13}$$

where we have used integration by parts since $\varphi_n^+(r)\varphi_m^-(r)$ vanish at the boundaries ($r = 0, r \to \infty$). Moreover, integration by parts also allows us to write

$$\langle \varphi_n^- | \varphi_m^- \rangle = \left(\frac{\lambdabar\omega}{\varepsilon+1}\right)^2 \langle \varphi_n^+ | \left[ -\frac{d^2}{dx^2} + \frac{\kappa(\kappa+1)}{x^2} \right] | \varphi_m^+ \rangle . \tag{14}$$

Using the differential equation, differential formula, and recursion relations of the Laguerre polynomials, we can show that the matrix representation of the Dirac operator (13) is tridiagonal if and only if $v_\pm = \pm(2\kappa+1) = 2|\kappa| \pm 1 = 2\ell+1$. Thus, the spinor basis in Eq. (11) and Eq. (12) could be written collectively for all $\kappa$ as follows

$$\chi_n(r,\varepsilon) = a_n x^\ell e^{-x/2} \begin{pmatrix} x L_n^{2\ell+1}(x) \\ \frac{1}{2}\frac{\lambdabar\omega}{\varepsilon+1}\left[ 2\kappa L_n^{2\ell+1}(x) - (n+2\ell+1)L_{n-1}^{2\ell+1}(x) + (n+1)L_{n+1}^{2\ell+1}(x) \right] \end{pmatrix}, \tag{15}$$

with $a_n = \sqrt{\omega \Gamma(n+1)/\Gamma(n+2\ell+2)}$. Using the identity formulas of the Laguerre polynomials we can rewrite this, for positive $\kappa$, to read

$$\chi_n(r,\varepsilon) = a_n x^\ell e^{-x/2} \begin{pmatrix} (n+2\ell+1)L_n^{2\ell}(x) - (n+1)L_{n+1}^{2\ell}(x) \\ \frac{1}{2}\frac{\lambdabar\omega}{\varepsilon+1}\left[ (n+2\ell+1)L_n^{2\ell}(x) + (n+1)L_{n+1}^{2\ell}(x) \right] \end{pmatrix}, \quad \kappa = 1, 2, 3, \ldots \tag{16a}$$

whereas, for negative $\kappa$, it simplifies to

$$\chi_n(r,\varepsilon) = a_n x^{\ell+1} e^{-x/2} \begin{pmatrix} L_n^{2\ell+2}(x) - L_{n-1}^{2\ell+2}(x) \\ -\frac{1}{2}\frac{\lambdabar\omega}{\varepsilon+1}\left[ L_n^{2\ell+2}(x) + L_{n-1}^{2\ell+2}(x) \right] \end{pmatrix}, \quad \kappa = -1, -2, -3, \ldots \tag{16b}$$

Now, after some simple, but somewhat lengthy, manipulations we obtain

$$\langle \phi_n^+ | \phi_m^+ \rangle = 2(n+\ell+1)\delta_{n,m} - \sqrt{n(n+2\ell+1)}\delta_{n-1,m} - \sqrt{(n+1)(n+2\ell+2)}\delta_{n+1,m} \tag{17a}$$

$$\langle \phi_n^- | \phi_m^- \rangle = \frac{1}{4}\left(\frac{\lambdabar\omega}{\varepsilon+1}\right)^2 \Big[ 2(n+\ell+1)\delta_{n,m} \\ + \sqrt{n(n+2\ell+1)}\delta_{n-1,m} + \sqrt{(n+1)(n+2\ell+2)}\delta_{n+1,m} \Big] \tag{17b}$$

Substituting these in Eq. (13) gives the following tridiagonal matrix representation of the reference wave operator

$$J_{nm}(\varepsilon) = \langle \chi_n | H_0 - \varepsilon | \chi_m \rangle = \frac{\lambdabar^2\omega^2}{\varepsilon+1}\left(\frac{2E}{\omega^2} + \frac{1}{4}\right) \times$$

$$\left[ -2\left(\frac{\frac{2E}{\omega^2} - \frac{1}{4}}{\frac{2E}{\omega^2} + \frac{1}{4}}\right)(n+\ell+1)\delta_{nm} + \sqrt{n(n+2\ell+1)}\delta_{n,m+1} + \sqrt{(n+1)(n+2\ell+2)}\delta_{n,m-1} \right], \tag{17c}$$

where $E(\varepsilon) \equiv (\varepsilon^2 - 1)/2\lambdabar^2 = \frac{1}{2}\mu^2$. In fact, in the nonrelativistic limit ($\lambdabar \to 0$) $E$ becomes the system's energy. Now, because the sine-like expansion coefficients $\{s_n\}_{n=0}^\infty$, satisfy the matrix-equivalent reference wave equation $\sum_{m=0}^{\infty} J_{n,m} s_m(\varepsilon) = 0$, then Eq. (17c) gives the following three-term recursion relation

$$2(n+\ell+1)(\cos\theta)s_n(\varepsilon) = \sqrt{n(n+2\ell+1)}s_{n-1}(\varepsilon) + \sqrt{(n+1)(n+2\ell+2)}s_{n+1}(\varepsilon), \quad (18)$$

where $\cos\theta = \left(\frac{2E}{\omega^2} - \frac{1}{4}\right) / \left(\frac{2E}{\omega^2} + \frac{1}{4}\right) = \left(\frac{\varepsilon^2-1}{\lambda^2\omega^2} - \frac{1}{4}\right) / \left(\frac{\varepsilon^2-1}{\lambda^2\omega^2} + \frac{1}{4}\right) = \left(\frac{\mu^2}{\omega^2} - \frac{1}{4}\right) / \left(\frac{\mu^2}{\omega^2} + \frac{1}{4}\right)$, and $0 < \theta \leq \pi$. Using the orthogonality relation of the Laguerre polynomials in the expansion $\phi_{reg}^+(r,\varepsilon) = \sum_n s_n(\varepsilon) \varphi_n^+(r)$, we can project out $s_n(\varepsilon)$ as

$$s_n(\varepsilon) = \frac{1}{\omega}\sqrt{\frac{\mu}{\omega}} a_n \int_0^\infty x^{\ell+\frac{1}{2}} e^{-x/2} L_n^{2\ell+1}(x) J_{\ell+\frac{1}{2}}\left(\frac{\mu}{\omega}x\right) dx, \quad (19)$$

where we have written the spherical Bessel function in terms of the regular Bessel function as $j_\nu(x) = \sqrt{\frac{\pi}{2x}} J_{\nu+\frac{1}{2}}(x)$. This integral could be evaluated analytically using the method proposed in [15] giving

$$s_n(\varepsilon) = \frac{a_n}{\omega\sqrt{2\pi}} 2^{\ell+1} \Gamma(\ell+1)(\sin\theta)^{\ell+1} C_n^{\ell+1}(\cos\theta) \quad (20)$$

where $C_n^\nu(z)$ is the Gegenbauer (ultra-spherical) polynomial [14]. Using the recursion relation of the $C_n^\nu(z)$, one can easily verify that $s_n(\varepsilon)$ as given by (20) satisfies the recursion relation (18) along with the initial condition $2(\ell+1)\cos\theta\, s_0 - \sqrt{2(\ell+1)}\, s_1 = 0$. Additionally, with the help of the differential equation of the Gegenbauer polynomials one can easily show that $s_n(\varepsilon)$ satisfies the following second order differential equation in the energy variable

$$\left[(1-y^2)\frac{d^2}{dy^2} - y\frac{d}{dy} - \frac{\ell(\ell+1)}{1-y^2} + (n+\ell+1)^2\right]s_n(\varepsilon) = 0. \quad (21)$$

Now, we look for a second independent solution to this second order differential equation. Let's call it $c_n(\varepsilon)$. Using the fact that $y = \pm 1$ are regular singularities of the equation, then Frobenius method dictates that the solution has the following form [16]

$$c_n(\varepsilon) = (1-y)^\alpha (1+y)^\beta f_n(\alpha,\beta;y), \quad (22)$$

where $\alpha$ and $\beta$ are real parameters such that $\beta$ is non-negative ($\beta \geq 0$) to prevent infrared divergence (at $\varepsilon = 1$ where $y = -1$). It should be obvious that the solution which simultaneously satisfies the recursion relation (18) and the differential equation (21) will be determined uniquely modulo an arbitrary overall factor, which is independent of $\varepsilon$ and $n$. That is, it will only depend on $\ell$ (i.e. depends only on $\kappa$) and we refer to it as $A_\ell$. Substituting (22) in place of $s_n(\varepsilon)$ in Eq. (21) shows that $f_n(\alpha,\beta;y)$ satisfies the same differential equation as the hyper-geometric function ${}_2F_1\left(a,b;c;\frac{1-y}{2}\right)$ [14] provided that

1) $c = 2\alpha + \frac{1}{2}$, $a = \alpha + \beta - (n+\ell+1)$, $b = \alpha + \beta + (n+\ell+1)$, (23a)

2) $\left(\beta - \frac{1}{4}\right)^2 + \left(\alpha - \frac{1}{4}\right)^2 = \frac{1}{2}\left(\ell + \frac{1}{2}\right)^2$, and (23b)

3) $\left(\beta - \frac{1}{4}\right)^2 = \left(\alpha - \frac{1}{4}\right)^2$ (23c)

The last equation (23c) gives two possibilities: $\alpha = \beta$ or $\alpha = -\beta + \frac{1}{2}$. For each one of these two possibilities, Eq. (23b) results in two alternative values for $\beta$: $\beta = \frac{1}{2}(\ell+1)$ or $\beta = -\frac{1}{2}\ell$. However, maintaining positivity of $\beta$, the latter is acceptable only for S-wave, where $\ell = 0$ (i.e., $\kappa = -1$), giving $\beta = 0$. Therefore, for S-wave we end up with two independent solutions corresponding to $\beta = \frac{1}{2}$ and $\beta = 0$. In the following subsections, we study the two cases given by (23c) separately.

### III.A. The case $\alpha = \beta$

We start by studying the solution that corresponds to $\beta = \frac{1}{2}(\ell+1)$ which is valid for all values of the spin-orbit quantum number $\kappa$ (i.e., for all $\ell$). In this case, Eqs. (23) give

$$\alpha = \beta = \tfrac{1}{2}(\ell+1),\ a = -n,\ b = n + 2\ell + 2,\ \text{and}\ c = \ell + \tfrac{3}{2}. \tag{24}$$

Substituting these parameters in (22) results in the expansion coefficients of the regular solution $\Phi_{reg}(r,\varepsilon)$ which we have already found in (20) and called it $s_n(\varepsilon)$. This could easily be seen by noting that $_2F_1\left(-n, n+2\ell+2; \ell+\tfrac{3}{2}; \tfrac{1-y}{2}\right)$ is proportional to $C_n^{\ell+1}(y)$ whereas $(1-y)^\alpha (1+y)^\beta = (1-y^2)^{\frac{1}{2}(\ell+1)} = (\sin\theta)^{\ell+1}$. However, for S-wave ($\ell = 0$) there exists another independent solution where,

$$\alpha = \beta = 0,\ a = -b = -n-1,\ \text{and}\ c = \tfrac{1}{2}, \tag{25}$$

corresponding to $_2F_1\left(-n-1, n+1; \tfrac{1}{2}; \tfrac{1-y}{2}\right)$, which is the Chebyshev polynomial of the first kind, $T_{n+1}(y)$ [14]. Therefore, the energy dependence of $c_n(\varepsilon)$ is now determined as $T_{n+1}(y)$. On the other hand, the $n$-dependent factor is determined from the recursion relation (18), which is satisfied by this $c_n(\varepsilon)$ with $\ell = 0$ but for $n \neq 0$. With the help of the recursion relation of the Chebyshev polynomials this factor is obtained as $1/\sqrt{n+1}$. Thus, this expansion coefficients of the reference wavefunction becomes

$$c_n(\varepsilon) = \tfrac{C}{\sqrt{n+1}} T_{n+1}(\cos\theta) = \tfrac{C}{\sqrt{n+1}} \cos(n+1)\theta, \tag{26}$$

where $C$ is a constant factor, which is independent of the energy $\varepsilon$ and the index $n$. Now, this solution satisfies the three term recursion relation (18) with $\ell = 0$, but not the initial relation (i.e., for $n = 0$). That's why we called it $c_n(\varepsilon)$ and not $s_n(\varepsilon)$. In fact, using the recursion relation of the Chebyshev polynomials one can easily verify that it satisfies the following inhomogeneous initial relation

$$2y c_0 = \sqrt{2}\, c_1 + C. \tag{27}$$

This is a crucial point. As stated at the beginning of this section, it means that the associated regularized wavefunction, $\Phi_{\cos}(r,\varepsilon) = \sum_n c_n(\varepsilon)\chi_n(r)$, does not solve the reference wave equation since $\sum_m J_{nm} c_m \neq 0$. However, Eq. (27) and the expression for $J_{nm}$ given by Eq. (17c) imply that

$$\sum_m J_{nm} c_m = -\frac{\hbar^2 \omega^2}{\varepsilon+1}\left(\frac{2E}{\omega^2} + \tfrac{1}{4}\right) C \delta_{n0} = -\frac{\hbar^2 \mu \omega C}{(\varepsilon+1)\sin\theta} \delta_{n0}. \tag{28}$$

This means that $\Phi_{\cos}$ for $\ell = 0$ solves the following regularized inhomogeneous wave equation

$$(H_0 - \varepsilon)|\Phi_{\cos}\rangle = -\frac{\hbar^2 \mu \omega C}{(\varepsilon+1)\sin\theta}|\tilde{\chi}_0\rangle, \tag{29}$$

The tridiagonal requirement of the orthogonal conjugate basis $\{\tilde{\chi}_n\}_{n=0}^{\infty}$ and that it should satisfy $\langle \chi_n | \tilde{\chi}_m \rangle = \langle \tilde{\chi}_n | \chi_m \rangle = \delta_{nm}$ dictate that the two components of $\tilde{\chi}_n$ must be a linear combination of $x^{|\kappa|} e^{-x/2} L_m^{2|\kappa|}(x)$, where $m = n, n \pm 1$ for $\pm \kappa > 0$. We obtain after some manipulations

$$\tilde{\chi}_n(r,\varepsilon) = \tfrac{1}{4\ell} a_n x^\ell e^{-x/2} \begin{pmatrix} (n+2\ell+1)L_n^{2\ell}(x) + (n+1)L_{n+1}^{2\ell}(x) \\ 2\frac{\varepsilon+1}{\hbar\omega}\left[(n+2\ell+1)L_n^{2\ell}(x) - (n+1)L_{n+1}^{2\ell}(x)\right] \end{pmatrix}, \quad \kappa > 0 \tag{30a}$$

$$\tilde{\chi}_n(r,\varepsilon) = \tfrac{1/4}{\ell+1} a_n x^{\ell+1} e^{-x/2} \begin{pmatrix} L_n^{2\ell+2}(x) + L_{n-1}^{2\ell+2}(x) \\ -2\frac{\varepsilon+1}{\hbar\omega}\left[L_n^{2\ell+2}(x) - L_{n-1}^{2\ell+2}(x)\right] \end{pmatrix}, \quad \kappa < 0 \tag{30b}$$

For $\ell = 0$, we obtain

$$\tilde{\chi}_0 = \tfrac{1}{4}\sqrt{\omega}\, x e^{-x/2} \begin{pmatrix} 1 \\ -2\frac{\varepsilon+1}{\hbar\omega} \end{pmatrix}. \tag{31}$$

Therefore, the right-hand side of Eq. (29) vanishes in the asymptotic region where $r \to \infty$. That is, asymptotically $\Phi_{\cos}$ satisfies the same reference wave equation as does $\Phi_{irr}$. Moreover, by equating the asymptotic behavior of $\Phi_{\cos}(r,\varepsilon)$ with that of $\Phi_{irr}(r,\varepsilon)$ and making use of the nonrelativistic limit, one can find the value of $C$, which is equal to $-\sqrt{2/\pi\omega}$. This will be confirmed in the following subsection.

### III.B. The case $\alpha = -\beta + \tfrac{1}{2}$

In this case, similar to the previous one, there are two alternative values for $\beta$: $\beta = \tfrac{1}{2}(\ell+1)$ and $\beta = -\tfrac{1}{2}\ell$. However, maintaining non-negativity of $\beta$, the second alternative is valid only for $\ell = 0$. Considering the first alternative, which is valid for all $\kappa$, Eqs. (23) give

$$\alpha = -\beta + \tfrac{1}{2} = -\tfrac{1}{2}\ell, \; a = -n-\ell-\tfrac{1}{2}, \; b = n+\ell+\tfrac{3}{2}, \text{ and } c = -\ell+\tfrac{1}{2}. \tag{32}$$

Thus, the resulting energy dependence of $c_n(\varepsilon)$ is given as

$$\left(\sin\tfrac{\theta}{2}\right)^{-\ell} \left(\cos\tfrac{\theta}{2}\right)^{\ell+1} {}_2F_1\left(-n-\ell-\tfrac{1}{2}, n+\ell+\tfrac{3}{2}; -\ell+\tfrac{1}{2}; \sin^2\tfrac{\theta}{2}\right). \tag{33}$$

This hypergeometric function is a non-terminating series because none of the first two arguments will ever be a negative integer. However, we can use the transformation [14],

$${}_2F_1(a,b;c;z) = (1-z)^{c-a-b}\, {}_2F_1(c-a, c-b; c; z), \tag{34}$$

to write it in the following alternative, but equivalent form

$$(\sin\theta)^{-\ell}\, {}_2F_1\left(-n-2\ell-1, n+1; -\ell+\tfrac{1}{2}; \sin^2\tfrac{\theta}{2}\right). \tag{35}$$

Now, this hypergeometric function is a finite polynomial of order $n+2\ell+1$ in $\sin^2\tfrac{\theta}{2}$. Requiring that this energy dependent function satisfy the three-term recursion relation (18) for $n \neq 0$ determines the $n$-dependence, which turns out to be proportional to $a_n$, giving the following

$$c_n(\varepsilon) = \tfrac{1}{\omega} A_\ell\, a_n (\sin\theta)^{-\ell}\, {}_2F_1\left(-n-2\ell-1, n+1; -\ell+\tfrac{1}{2}; \sin^2\tfrac{\theta}{2}\right), \tag{36}$$

where $A_\ell$ is an overall constant factor, which is independent of the energy $\varepsilon$ and the index $n$. It could be evaluated by equating the asymptotic behavior of $\Phi_{\cos}(r,\varepsilon)$ with that of $\Phi_{irr}(r,\varepsilon)$. However, a simpler and more straight-forward method is by taking the non-relativistic limit ($\lambdabar \to 0$, $\varepsilon \to 1+\lambdabar^2 E$) of this result. Doing so gives the nonrelativistic cosine-like expansion coefficients $c_n(E)$ in the Laguerre basis for the case corresponding to $\nu_+ = 2\ell+1$ (i.e., $\kappa > 0$) [7]. Paying special attention to the choice of normalization of the regular solution and basis in [7] that differs from ours, we obtain $A_\ell = -\frac{1}{\pi} 2^{\ell+\frac{1}{2}} \Gamma(\ell+\frac{1}{2})$. One can verify that $c_n(\varepsilon)$ satisfies the three-term recursion relation (18) but not the initial relation (when $n = 0$). Instead, it satisfies the following inhomogeneous initial relation

$$2(\ell+1)y c_0 = \sqrt{2\ell+2}\, c_1 + A_\ell \sqrt{\frac{2\ell+1}{\omega \Gamma(2\ell+1)}} (\sin\theta)^{-\ell}. \tag{37}$$

Therefore, the corresponding wavefunction does not satisfy the reference wave equation but an inhomogeneous one that reads

$$(H_0 - \varepsilon)|\Phi_{\cos}\rangle = -A_\ell \sqrt{\frac{\omega(2\ell+1)}{\Gamma(2\ell+1)}} \frac{\lambdabar^2 \mu}{\varepsilon+1} (\sin\theta)^{-\ell-1} |\tilde{\chi}_0\rangle, \tag{38}$$

where $\tilde{\chi}_0(r,\varepsilon)$ is now given for all $\kappa$ by Eq. (30) as $\tilde{\chi}_0 \sim x^{|\kappa|} e^{-x/2}(a_\pm + b_\pm x)\binom{1}{\pm 2\frac{\varepsilon+1}{\lambdabar\omega}}$, where $a_\pm$ and $b_\pm$ are $\kappa$-dependent constants. It is not difficult to verify that the S-wave solution obtained above in (26) is a special case of (36) with $\ell=0$ ($\kappa=-1$) and $C = A_0/\sqrt{\omega}$. Similar to the previous case ($\alpha=\beta$), we also find another independent special solution for S-wave ($\ell=0$) where,

$$\alpha = \tfrac{1}{2},\ \beta = 0,\ a = -n-\tfrac{1}{2},\ b = n+\tfrac{3}{2},\ \text{and}\ c = \tfrac{3}{2}, \tag{39}$$

corresponding to $(1-y)^{\frac{1}{2}} {}_2F_1\left(-n-\tfrac{1}{2}, n+\tfrac{3}{2}; \tfrac{3}{2}; \tfrac{1-y}{2}\right)$. Using the transformation (34) this could be rewritten as $(\sin\theta)\, {}_2F_1\left(-n, n+2; \tfrac{3}{2}; \tfrac{1-y}{2}\right)$. Alternatively, we could write it as $\frac{\sin\theta}{n+1} U_n(y)$, where $U_n(y)$ is the Chebyshev polynomial of the second kind [14]. Moreover, requiring that this solution satisfy the three-term recursion relation (18) for $\ell=0$ gives the following expansion coefficients of the reference wavefunction

$$s_n(\varepsilon) = \frac{B \sin\theta}{\sqrt{n+1}} U_n(\cos\theta) = \frac{B}{\sqrt{n+1}} \sin(n+1)\theta, \tag{40}$$

where $B$ is a constant factor, which is independent of the energy $\varepsilon$ and the index $n$. One can easily verify that this solution satisfies the initial relation ($n = 0$) of (18) with $\ell=0$ ($\kappa=-1$). That's why it was referred to as $s_n(\varepsilon)$. In fact, one can easily show that this solution is a special case of that in (20) with $\ell=0$ and $B = \sqrt{2/\pi\omega}$.

In summary, we have obtained two independent solutions of the reference problem (the 3D free Dirac equation with spherical symmetry) which are valid asymptotically. These are $\Phi_{reg} = \Phi_{\sin} = \sum_n s_n \chi_n$ and $\Phi_{\cos} = \sum_n c_n \chi_n$, where $s_n$ and $c_n$ are given by Eq. (20) and Eq. (36), respectively. Now, since the potential functions and effective mass are assumed finite in range, then these two solutions are also related to the total spinor wave function in the asymptotic region. We can write

$$\lim_{r\to\infty} \Psi(r,\varepsilon) = \Phi_-(r,\varepsilon) + e^{2i\delta(\varepsilon)} \Phi_+(r,\varepsilon), \tag{41}$$

–11–

where $\Phi_\pm(r,\varepsilon) = \Phi_{\cos}(r,\varepsilon) \pm i\Phi_{\sin}(r,\varepsilon)$ and $\delta(\varepsilon)$ is the energy-dependent phase shift for a given effective mass distribution $S(r)$ and potential functions $V(r)$ and $W(r)$.

Contrary to the asymptotic reference solutions, which are obtained above analytically, the phase shift will be obtained numerically. One can calculate $\delta(\varepsilon)$ using any convenient approach based on the chosen scattering method. In the following section, we use the relativistic J-matrix method [10,11] to obtain $\delta(\varepsilon)$.

## IV. RELATIVISTIC J-MATRIX SCATTERING MATRIX: DYNAMICS

The vector and pseudo scalar potentials, $V(r)$ and $W(r)$, are assumed short-range such that they are represented accurately by their matrix elements in a finite set of $L^2$ basis functions $\{\zeta_n(r)\}_{n=0}^{N-1}$, for some large enough integer $N$. The relativistic effective mass, $S(r)$, is also localized accurately in the same finite basis. In contrast to the kinematic basis (15) this finite basis, which is evidently a two-component spinor basis, should be energy independent. Moreover, it is required to sample the objects defined in the interior scattering region (i.e., objects like $V$, $W$ and $S$ that are short-range) to a high degree of accuracy. That is, the following finite $N \times N$ matrix representation of the potential $V(r)$

$$V_{nm} = \begin{cases} \langle \zeta_n | V | \zeta_m \rangle & ; n,m = 0,1,..,N-1 \\ 0 & ; n,m \geq N \end{cases} \quad (42)$$

is an approximation that is accurate enough for some given integer $N$. This should also hold true for the matrix representation of both $W(r)$ and $S(r)$.

We write the total spinor wavefunction $\Psi(r,\varepsilon)$, which is a solution of Eq. (2), as the sum of two parts. One belongs to the inner (scattering) region in function space, which is written in terms of the energy-independent basis $\{\zeta_n(r)\}_{n=0}^{N-1}$. The other belongs to the outer (asymptotic) region, which is written in terms of the energy-dependent basis $\{\chi_n(r,\varepsilon)\}_{n=N}^{\infty}$. Now, if we define $\mathbb{P}(\varepsilon)$ as the projection operator in the space spanned by the total spinor wave function $|\Psi\rangle$. Then, it will be the sum of two parts, $\mathbb{P}(\varepsilon) = \mathbb{P}_{in} + \mathbb{P}_{out}(\varepsilon)$. The "inside" projection operator, $\mathbb{P}_{in} = \sum_{n=0}^{N-1} |\zeta_n\rangle\langle\tilde{\zeta}_n|$, is of dimension $N$, whereas the "outside" projection operator, $\mathbb{P}_{out}(\varepsilon) = \sum_{n=N}^{\infty} |\chi_n\rangle\langle\tilde{\chi}_n|$, is energy-dependent and of infinite dimension. The basis elements with the tilde symbol on top are those that span the orthogonal conjugate subspace (e.g., $\langle\zeta_n|\tilde{\zeta}_m\rangle = \langle\tilde{\zeta}_n|\zeta_m\rangle = \delta_{nm}$). Moreover, the two subspaces do not overlap. That is, $\langle\tilde{\zeta}_n|\chi_m\rangle = \langle\tilde{\chi}_m|\zeta_n\rangle = 0$ for all $n = 0,1,...,N-1$ and $m \geq N$. Consequently, we can write the total spinor wavefunction as

$$|\Psi(r,\varepsilon)\rangle = \sum_{n=0}^{N-1} p_n(\varepsilon)|\zeta_n(r)\rangle + \sum_{n=N}^{\infty} q_n(\varepsilon)|\chi_n(r,\varepsilon)\rangle, \quad (43)$$

where $p_n = \langle\tilde{\zeta}_n|\Psi\rangle$ and $q_n = \langle\tilde{\chi}_n|\Psi\rangle$. The J-matrix representation of the total wave function in the asymptotic region is shown in Eq. (41) and could be rewritten as

$$\mathbb{P}_{out}|\Psi\rangle = \sum_{n=N}^{\infty}\left[(c_n - \mathrm{i}\, s_n) + e^{2\mathrm{i}\delta}(c_n + \mathrm{i}\, s_n)\right]|\chi_n\rangle. \tag{44}$$

Therefore, $q_n = (c_n - \mathrm{i}\, s_n) + e^{2\mathrm{i}\delta}(c_n + \mathrm{i}\, s_n)$ and the solution of the full scattering problem is obtained if we could calculate the $N+1$ unknowns: the coefficients $\{p_n\}_{n=0}^{N-1}$ and phase shift $\delta$. The J-matrix method provides a platform for such a solution [7].

Evidently, Eq. (43) indicates that the basis for the total solution space of the problem splits into an inner and outer components that is collectively written as
$$(\zeta_0, \zeta_1, ..., \zeta_{N-1}, \chi_N, \chi_{N+1}, \chi_{N+2}, ...), \tag{45}$$
where $\zeta_n = \begin{pmatrix}\xi_n^+\\ \xi_n^-\end{pmatrix}$. The requirement that the reference wave operator $J = H_0 - \varepsilon$ maintains its tridiagonal structure outside the inner scattering region in function space (i.e., $\langle\zeta_n|J|\chi_m\rangle = 0$ for $n = 0,1,..,N-1$ and $m \geq N$, except that $\langle\zeta_{N-1}|J|\chi_N\rangle = \langle\chi_N|J|\zeta_{N-1}\rangle \neq 0$) dictates that $\xi_n^+(r) = \varphi_n^+(r)$ and gives $\langle\zeta_n|J|\chi_m\rangle = \langle\chi_n|J|\chi_m\rangle$. We propose an expression for the lower component $\xi_n^-$ that results in a tridiagonal overlap matrix $\langle\zeta_n|\zeta_m\rangle$. This, of course, is not necessary but has the advantage of making the calculation of the matrix elements of the potentials, effective mass, and reference Hamiltonian simpler. These calculations could be carried out using Gauss quadrature integral approximation [17]. Moreover, the expression of the Green's function associated with the finite $N\times N$ total Hamiltonian, which enters in the computation of the scattering matrix, becomes also neat and simple. The proposed ansatz for $\xi_n^-$ is the energy-independent "kinetic-balance-like" relation $\xi_n^- = \frac{\lambdabar\omega}{\rho}\left(\frac{\kappa}{x} + \frac{d}{dx}\right)\xi_n^+$, where $\rho$ is a (second) real basis parameter [13]. Therefore, we can rewrite $\xi_n^-$ as $\xi_n^- = \frac{\varepsilon+1}{\rho}\varphi_n^-$ and thus the two-component spinor basis element $\zeta_n(r)$ now becomes

$$\zeta_n(r) = a_n\, x^\ell e^{-x/2}\begin{pmatrix}(n+2\ell+1)L_n^{2\ell}(x) - (n+1)L_{n+1}^{2\ell}(x)\\ \frac{\lambdabar\omega}{2\rho}\left[(n+2\ell+1)L_n^{2\ell}(x) + (n+1)L_{n+1}^{2\ell}(x)\right]\end{pmatrix}, \; \kappa > 0 \tag{46a}$$

$$\zeta_n(r) = a_n\, x^{\ell+1} e^{-x/2}\begin{pmatrix}L_n^{2\ell+2}(x) - L_{n-1}^{2\ell+2}(x)\\ -\frac{\lambdabar\omega}{2\rho}\left[L_n^{2\ell+2}(x) + L_{n-1}^{2\ell+2}(x)\right]\end{pmatrix}, \; \kappa < 0 \tag{46b}$$

The resulting overlap matrix is tridiagonal and reads as follows
$$\langle\zeta_n|\zeta_m\rangle = 2\sigma_+(n+\ell+1)\delta_{nm}$$
$$-\sigma_-\left[\sqrt{n(n+2\ell+1)}\,\delta_{n,m+1} + \sqrt{(n+1)(n+2\ell+2)}\,\delta_{n,m-1}\right], \tag{47}$$

where $\sigma_\pm = 1 \pm (\lambdabar\omega/2\rho)^2$. Additionally, the matrix representation of the reference Hamiltonian (the free Dirac Hamiltonian $H_0$) in this inner basis is tridiagonal and is written as follows
$$(H_0)_{n,m} = \langle\zeta_n|H_0|\zeta_m\rangle = 2\rho_+(n+\ell+1)\delta_{n,m}$$
$$-\rho_-\left[\sqrt{n(n+2\ell+1)}\,\delta_{n,m+1} + \sqrt{(n+1)(n+2\ell+2)}\,\delta_{n,m-1}\right], \tag{48}$$

where $\rho_\pm = 1 \pm (2\rho-1)(\lambdabar\omega/2\rho)^2$. Taking the nonrelativistic limit ($\lambdabar \to 0$) gives $H_0 \to 1 + \lambdabar^2 \hat{H}_0$, where $\hat{H}_0$ is the nonrelativistic reference Hamiltonian. In matrix notation, $(H_0)_{nm} \to \langle \zeta_n | \zeta_m \rangle + \lambdabar^2 (\hat{H}_0)_{nm} = \langle \xi_n^+ | \xi_m^+ \rangle + \lambdabar^2 [\langle \xi_n^- | \xi_m^- \rangle \lambdabar^{-2}] + \lambdabar^2 (\hat{H}_0)_{nm}$. For the case $v_+ = 2\ell+1$ (i.e., $\kappa > 0$), $(\hat{H}_0)_{nm}$ become the matrix elements of the nonrelativistic J-matrix representation of the reference Hamiltonian in the Laguerre basis [7]. This identification results in the choice $\rho = 2$. However, we will keep the value of $\rho$ at present arbitrary.

Now, the matrix representation of the 2×2 short range mass-potential function, $\mathcal{V} = \lambdabar \begin{pmatrix} \lambdabar V_+ & W \\ W & \lambdabar V_- \end{pmatrix}$, in the same inner subspace is as follows

$$\mathcal{V}_{nm} = \langle \zeta_n | \mathcal{V} | \zeta_m \rangle = \lambdabar^2 \left[ \langle \xi_n^+ | V_+ | \xi_m^+ \rangle + \langle \xi_n^- | V_- | \xi_m^- \rangle \right] + \lambdabar \left[ \langle \xi_n^+ | W | \xi_m^- \rangle + \langle \xi_n^- | W | \xi_m^+ \rangle \right], \quad (49)$$

where $n, m = 0, 1, .., N-1$. Let's define the $(n,m)$ sampling element of a real radial function $F(r)$, which is not necessarily differentiable, by the Laguerre polynomials as the value of the integral

$$F_{nm}^v = \sqrt{\frac{\Gamma(n+1)\Gamma(m+1)}{\Gamma(n+v+1)\Gamma(m+v+1)}} \int_0^\infty x^v e^{-x} L_n^v(x) F(x/\omega) L_m^v(x) \, dx. \quad (50)$$

Then substituting the components $\xi_n^\pm$ from Eq. (46) into (49) and using this integral definition, we obtain for $\pm \kappa > 0$

$$\langle \xi_n^+ | V_+ | \xi_m^+ \rangle = (U_+)_{nm}^{2\ell+1}, \quad (51a)$$

$$\langle \xi_n^+ | W | \xi_m^- \rangle + \langle \xi_n^- | W | \xi_m^+ \rangle = \pm \frac{\lambdabar \omega}{\rho} \left[ \sqrt{(n_\pm + 2|\kappa|)(m_\pm + 2|\kappa|)} \, W_{n,m}^{2|\kappa|} - \sqrt{n_\pm m_\pm} \, W_{n\pm 1, m\pm 1}^{2|\kappa|} \right] \quad (51b)$$

$$\langle \xi_n^- | V_- | \xi_m^- \rangle = \left(\frac{\lambdabar \omega}{2\rho}\right)^2 \left[ \sqrt{(n_\pm + 2|\kappa|)(m_\pm + 2|\kappa|)} (V_-)_{n,m}^{2|\kappa|} + \sqrt{n_\pm m_\pm} (V_-)_{n\pm 1, m\pm 1}^{2|\kappa|} \right. \\ \left. + \sqrt{m_\pm (n_\pm + 2|\kappa|)} (V_-)_{n, m\pm 1}^{2|\kappa|} + \sqrt{n_\pm (m_\pm + 2|\kappa|)} (V_-)_{m, n\pm 1}^{2|\kappa|} \right] \quad (51c)$$

where $n_+ = n+1$ and $n_- = n$ for $\kappa > 0$ and $\kappa < 0$, respectively. The radial function $U_+$ in Eq. (51a) is defined as $U_+(r) = (\omega r) V_+(r)$. These formulas show that the contribution of $V_+(r)$ and $W(r)$ to the matrix elements of the total Hamiltonian is second order in the relativistic parameter (the Compton wavelength) $\lambdabar$, whereas for $V_-(r)$ it is fourth order. For an integer $K \geq N$, we use Gauss quadrature [17] to give the following approximation for the integral (50)

$$F_{nm}^v \cong \sum_{k=0}^{K-1} \Lambda_{nk}^v \Lambda_{mk}^v F(\eta_k^v / \omega), \quad (52)$$

where $\{\Lambda_{nk}^v\}_{n=0}^{K-1}$ is the normalized eigenvector associated with the eigenvalue $\eta_k^v$ of the $K \times K$ tridiagonal symmetric matrix of the quadrature associated with the Laguerre polynomials $\{L_n^v\}$. That is, the matrix whose elements are $(2n+v+1)\delta_{n,m} + \sqrt{n(n+v)} \, \delta_{n,m+1} + \sqrt{(n+1)(n+v+1)} \, \delta_{n,m-1}$ for $n, m = 0, 1, 2, .., K-1$. Thus, the reference Hamiltonian, $H_0$, is fully accounted for in the whole representation space (45) as given by Eq. (48) and in Eq. (17c). However, the mass-potential matrix $\mathcal{V}$ is approximated by its representation in a subset of the basis. That is,

$$H_{nm} \cong \begin{cases} (H_0)_{nm} + \mathcal{V}_{nm} & ; \quad n,m \leq N-1 \\ (H_0)_{nm} & ; \quad n,m > N-1 \end{cases}. \tag{53}$$

This representation is the fundamental underlying structure of certain algebraic scattering methods, such as the R-matrix [9] and J-matrix methods [7]. One may confine investigation to the finite $N \times N$ matrix representation of the potential $\mathcal{V}$ and the reference Hamiltonian $H_0$. However, taking into account the full contribution of the reference Hamiltonian, as will be done in this work, should result in a substantial improvement on the accuracy of the results. Now, if we define the "J-matrix coefficients"

$$T_n(\varepsilon) = \frac{c_n(\varepsilon) - \mathrm{i}\, s_n(\varepsilon)}{c_n(\varepsilon) + \mathrm{i}\, s_n(\varepsilon)}, \quad R_n^{\pm}(\varepsilon) = \frac{c_n(\varepsilon) \pm \mathrm{i}\, s_n(\varepsilon)}{c_{n-1}(\varepsilon) \pm \mathrm{i}\, s_{n-1}(\varepsilon)}, \tag{54}$$

then the J-matrix method gives the following expression for the $N^{\text{th}}$ order relativistic S-matrix [11]

$$S^{(N)}(\varepsilon) = T_{N-1}(\varepsilon) \frac{1 + G_{N-1,N-1}(\varepsilon) J_{N-1,N}(\varepsilon) R_N^{-}(\varepsilon)}{1 + G_{N-1,N-1}(\varepsilon) J_{N-1,N}(\varepsilon) R_N^{+}(\varepsilon)} = e^{2\mathrm{i}\delta(\varepsilon)}, \tag{55}$$

where $G_{N-1,N-1}(\varepsilon)$ is the ($N-1,N-1$) component of the finite Green's matrix function which contains the dynamics by incorporating the contribution of the $2 \times 2$ short range mass-potential matrix $\mathcal{V}$,

$$G_{N-1,N-1}(\varepsilon) = \left\langle \tilde{\zeta}_{N-1} \left| (H_0 + \mathcal{V} - \varepsilon \otimes I)^{-1} \right| \tilde{\zeta}_{N-1} \right\rangle, \tag{56}$$

where $I$ is the $2 \times 2$ unit matrix. Due to the fact that the basis of the $L^2$ subspace, $\{\zeta_n\}_{n=0}^{N-1}$, is trithogonal, the finite Green's functions should be calculated as shown in Appendix B of the first reference in [11]. Using the recursion relation (18) satisfied by the coefficients $c_n$ and $s_n$ for $n \geq 1$ we can evaluate $T_{N-1}$ and $R_N^{\pm}$ recursively starting with $T_0$ and $R_1^{\pm}$ as

$$R_{n+1}^{\text{sgn}}(\varepsilon) = \frac{1}{\sqrt{(n+1)(n+2\ell+2)}} \left[ 2(n+\ell+1)(\cos\theta) - \sqrt{n(n+2\ell+1)} \Big/ R_n^{\text{sgn}}(\varepsilon) \right], \tag{57}$$

$$T_n(\varepsilon) = T_{n-1}(\varepsilon) \left[ R_n^{-}(\varepsilon) \Big/ R_n^{+}(\varepsilon) \right]. \tag{58}$$

for $n = 1, 2, \ldots, N-1$ and where the superscript "sgn" on the J-matrix coefficient $R$ stands for the $\pm$ sign.

In the following section we calculate the scattering phase shift for a given configuration specified by the choice of potential functions $V(r)$ and $W(r)$ and for an effective mass distribution $S(r)$. Taking $\rho = 2$, the results will be shown to be independent of the spinor basis parameter $\omega$ once we reach the plateau of computational stability. The computer code (RMJ-07.01) used in the calculation was developed using the Mathcad® software package, and is available upon request from the corresponding author.

## V. RESULTS AND DISCUSSION

To illustrate the utility and accuracy of our development, we choose the following mass-potential example in a single channel configuration

$$V_{\pm}(r) = V_{\pm}^{0} h(r), \quad W(r) = 0, \tag{59}$$

---

® Mathcad is a software package developed by Mathsoft for general-purpose mathematical computations.

where $V_\pm^0 = V_0 \pm S_0$ are real parameters and we consider the following radial potential function parameterized by the length parameter $r_0$:

$$h(r) = e^{-(r-r_0)^2/4} - 2e^{-r^2/5} \tag{60}$$

The calculation will be carried out in the Laguerre basis given above in Eq. (46a) and Eq. (46b). Alternatively, the same calculation could as well be done in the oscillator basis given by Eq. (A12a) and Eq. (A12b) in the Appendix below. We start by considering the constant mass case where $M(r) = m_0$ (i.e., $S_0 = 0$). Figure 1 is a plot of $\left|1 - S^{(N)}(\varepsilon)\right|$ for S-wave scattering as a function of the energy variable $E(\varepsilon)$ defined below Eq. (17c). The physical parameters were taken as $V_0 = 3.0$, and $r_0 = 3.0$ (all in atomic units). Moreover, we took a basis size $N = 50$ and $\omega = 10$ (a.u.). The solid curve is the relativistic result corresponding to $\lambdabar = 0.2$ (a.u.) whereas the dashed curve corresponds to the nonrelativistic limit where $\lambdabar = 0.001$ (a.u.). That is, for the nonrelativistic limit the speed of light was ascribed a value that is 200 times larger than $c$. Now, as stated in the Introduction, nonrelativistic position-dependent mass systems have ordering ambiguity that makes their solution not unique. Consequently, a consistency check of our results (in the limit) could only be made by comparison with those of constant mass. Figure 2 is the result of the non-relativistic calculation for the same problem with constant mass, which was performed using the standard nonrelativistic J-matrix method in the Laguerre basis with the same parameters. The agreement between Fig. 2 and the dashed curve in Fig. 1 is obvious. As expected, no significant differences exist at lower energies between the relativistic and nonrelativistic results. However, it is clear that the sharp resonance at the higher energy $E = 2.2$ (a.u.) in Fig. 1 becomes, in the nonrelativistic limit, less pronounced (broader) and shifts a little towards higher energy at $E = 2.3$ (a.u.). Moreover, the broad resonance near $E \sim 3.2$ (a.u.) also becomes harder to resolve in the nonrelativistic limit. In fact, a more precise calculation of the resonance structure for this potential could be performed using, for example, the complex scaling method [18]. Such calculation gives the relativistic resonances: $E = 2.2098 - i\, 0.0548$ and $E = 3.5786 - i\, 0.9663$ (a.u.). On the other hand, the nonrelativistic resonance energies are $E = 2.3272 - i\, 0.10486$ and $E = 3.5482 - i\, 1.2210$ (a.u.). Figure 3 gives the integrated phase shift $\delta(E)$ as a function of the energy for the relativistic (solid curve) and nonrelativistic (dashed curve) case. Now, we turn to the most relevant contribution of our work: the effect of spatial variation of the mass on the relativistic scattering matrix. Figure 4 and Fig. 5 are reproductions of Fig. 1 and Fig. 3 for the same system but for non-vanishing $S_0$. We took $S_0 = 1.0$ (a.u.) making $m(r) = 1 + \lambdabar^2 h(r)$. One of the obvious effects is the extreme sharpening and downshift of the resonance at $E = 2.1$ a.u. (relativistic) and $E = 2.3$ a.u. (nonrelativistic). In fact, the relativistic resonance structure of the system with this position-dependent mass becomes $E = 2.1211 - i\, 0.003185$ and $E = 3.9429 - i\, 0.4512$ (a.u.). The nonrelativistic limit of this structure is $E = 2.30333 - i\, 0.009111$ and $E = 3.9270 - i\, 0.6693$ (a.u.). Nonetheless, the full effect on the S-matrix for all energies (in the range of interest) is shown in Fig. 4 and Fig. 5.

Finally, we like to remark that the extension of the above development to multi-channel scattering is straightforward, which could be carried out following the same scheme as in the classic J-matrix method [19].

## ACKNOWLEDGMENTS

The Authors acknowledge the support of King Fahd University of Petroleum and Minerals via project FT-2006/05. Fruitful discussions with H. A. Yamani are highly appreciated. ADA is grateful to Amjad A. Al-Haidari (AUS) for providing literature resources in support of this work.

## APPENDIX
## J-MATRIX REPRESENTATION IN THE OSCILLATOR BASIS

In the upper component of this spinor basis given by Eq. (11), we replace $x$ by $z = x^2 = (\omega r)^2$. The lower spinor component is calculated using the kinetic balance relation, which now reads $\varphi_n^- = 2\frac{\hbar\omega}{\varepsilon+1} x \left(\frac{\kappa/2}{z} + \frac{d}{dz}\right)\varphi_n^+$. Thus, we obtain the energy-dependent spinor basis for the outer function space ($n \geq N$) as

$$\chi_n(r,\varepsilon) = a_n^\pm x^{\nu_\pm - \frac{1}{2}} e^{-x^2/2} \begin{pmatrix} x L_n^{\nu_\pm}(x^2) \\ \frac{\hbar\omega}{\varepsilon+1}\left[\left(\kappa - \frac{1}{2}\right)L_n^{\nu_\pm}(x^2) - (n+\nu_\pm)L_{n-1}^{\nu_\pm}(x^2) + (n+1)L_{n+1}^{\nu_\pm}(x^2)\right] \end{pmatrix}, \quad \text{(A1)}$$

corresponding to $\pm\kappa > 0$. Equivalently, we can write this as

$$\chi_n(r,\varepsilon) = a_n^\pm x^{\nu_\pm - \frac{1}{2}} e^{-x^2/2} \begin{pmatrix} x L_n^{\nu_\pm}(x^2) \\ \frac{\hbar\omega}{\varepsilon+1}\left\{\left(\kappa + \nu_\pm + \frac{1}{2}\right)L_n^{\nu_\pm}(x^2) - x^2\left[L_n^{\nu_\pm+1}(x^2) + L_{n-1}^{\nu_\pm+1}(x^2)\right]\right\} \end{pmatrix}. \quad \text{(A2)}$$

Due to the factor $\frac{1}{2\omega}$ in the integration measure in this basis, which is $\int_0^\infty dr = \frac{1}{2\omega}\int_0^\infty \frac{1}{\sqrt{z}}dz$, the normalization constant is taken as $a_n^\pm = \sqrt{2\omega\Gamma(n+1)/\Gamma(n+\nu_\pm+1)}$. Now, requiring that the matrix representation of the free Dirac operator in this basis be tridiagonal dictates that $\nu_\pm = \pm\left(\kappa + \frac{1}{2}\right) = |\kappa| \pm \frac{1}{2} = \ell + \frac{1}{2}$. Using the recursion relation of the Laguerre polynomials and their orthogonality relation, we obtain the following

$$J_{n,m}(\varepsilon) = \langle \chi_n | H_0 - \varepsilon | \chi_m \rangle = \frac{\hbar^2\omega^2}{\varepsilon+1}\left[-\frac{\varepsilon^2-1}{\hbar^2\omega^2}\delta_{n,m} + \left(2n+\ell+\tfrac{3}{2}\right)\delta_{n,m}\right.$$
$$\left. + \sqrt{n\left(n+\ell+\tfrac{1}{2}\right)}\delta_{n,m+1} + \sqrt{(n+1)\left(n+\ell+\tfrac{3}{2}\right)}\delta_{n,m-1}\right], \quad \text{(A3)}$$

Consequently, the reference wave equation becomes equivalent to the following three term recursion relation for the expansion coefficients of the wavefunction

$$y\, s_n(\varepsilon) = \left(2n+\ell+\tfrac{3}{2}\right)s_n(\varepsilon) + \sqrt{n\left(n+\ell+\tfrac{1}{2}\right)}s_{n-1}(\varepsilon) + \sqrt{(n+1)\left(n+\ell+\tfrac{3}{2}\right)}s_{n+1}(\varepsilon), \quad \text{(A4)}$$

where $y = \left(\frac{\mu}{\omega}\right)^2 = \frac{\varepsilon^2-1}{\hbar^2\omega^2} = \frac{2E}{\omega^2}$. Using the orthogonality relation of the Laguerre polynomials in the expansion $\phi_{reg}^+(r,\varepsilon) = \sum_n s_n(\varepsilon)\varphi_n^+(r)$, we can project out $s_n(\varepsilon)$ as

$$s_n(\varepsilon) = \frac{1}{\omega}\sqrt{\frac{\mu}{\omega}} a_n \int_0^\infty x^{\ell+\frac{3}{2}} e^{-x^2/2} L_n^{\ell+\frac{1}{2}}(x^2) J_{\ell+\frac{1}{2}}\left(\frac{\mu}{\omega}x\right) dx, \quad \text{(A5)}$$

where $a_n = \sqrt{2\omega\Gamma(n+1)/\Gamma\left(n+\ell+\tfrac{3}{2}\right)}$. Contrary to the integral (19) above, which is not found in all known tables of integrals, this integral is available in most of them giving

$$s_n(\varepsilon) = (-1)^n \frac{1}{\omega} a_n \left(\frac{\mu}{\omega}\right)^{\ell+1} e^{-\mu^2/2\omega^2} L_n^{\ell+\frac{1}{2}}\left(\mu^2/\omega^2\right), \quad \text{(A6)}$$

Writing $s_n(\varepsilon) \sim y^{\frac{1}{2}(\ell+1)} e^{-y/2} L_n^{\ell+\frac{1}{2}}(y)$ and using the differential equation of the Laguerre polynomials, we can show that $s_n(\varepsilon)$ satisfies the following second order differential equation in $y$

$$\left[ y \frac{d^2}{dy^2} + \frac{1}{2}\frac{d}{dy} - \frac{\ell(\ell+1)}{4y} - \frac{1}{4}y + \frac{1}{2}\left(2n+\ell+\tfrac{3}{2}\right) \right] s_n(\varepsilon) = 0. \tag{A7}$$

A second independent solution of this equation could be obtained by writing $s_n(\varepsilon) = y^\alpha e^{-\beta y} f_n(\alpha,\beta;y)$, where $\beta > 0$. The resulting differential equation for $f_n(\alpha,\beta;y)$ will be the same as that of the confluent hyper-geometric function ${}_1F_1(a;c;y)$ [14] provided that

1) $\beta = 1/2$ and $c = 2\alpha + \tfrac{1}{2}$, \hfill (A8a)

2) $\left(\alpha - \tfrac{1}{4}\right)^2 = \tfrac{1}{4}\left(\ell + \tfrac{1}{2}\right)^2$, and \hfill (A8b)

3) $a = -n + \alpha - \tfrac{1}{2}(\ell+1)$. \hfill (A8c)

The case where $\alpha = \tfrac{1}{2}(\ell+1)$ reproduces $s_n(\varepsilon)$. However, $\alpha = -\tfrac{1}{2}\ell$ gives the new independent solution: $c_n(\varepsilon) \sim y^{-\frac{1}{2}\ell} e^{-y/2} {}_1F_1\left(-n-\ell-\tfrac{1}{2}; -\ell+\tfrac{1}{2}; y\right)$. Requiring that this solution satisfies the three-term recursion relation (A4) (for $n \neq 0$) and using the following relation of the confluent hyper-geometric series [14]

$$z\,{}_1F_1(a;c;z) = (c-2a)\,{}_1F_1(a;c;z) + a\,{}_1F_1(a+1;c;z) + (a-c)\,{}_1F_1(a-1;c;z), \tag{A9}$$

will determine the $n$-dependent factor in $c_n(\varepsilon)$ as $(-1)^n a_n$. Thus, we are left with an overall factor that is independent of the energy and the index $n$ (designated as $A_\ell$) giving

$$c_n(\varepsilon) = \tfrac{1}{\omega} A_\ell (-1)^n a_n (\mu/\omega)^{-\ell} e^{-\mu^2/2\omega^2} {}_1F_1\left(-n-\ell-\tfrac{1}{2}; -\ell+\tfrac{1}{2}; \mu^2/\omega^2\right) \tag{A10}$$

Taking the nonrelativistic limit ($\lambdabar \to 0$) [7] gives $A_\ell = \tfrac{\sqrt{2}}{\pi}\Gamma\left(\ell+\tfrac{1}{2}\right)$. The confluent hyper-geometric series in (A10) is non-terminating since the negative of the first argument is half of odd integer. Therefore, for large values of $N$ and higher energies, $\varepsilon$, numerical instability could, in principle, occur when trying to evaluate $c_N(\varepsilon)$ needed for the calculation of the scattering matrix. This is due to the fact that in such circumstances [as shown in Eq. (A10)] we would be multiplying a very small number coming from the decaying exponential $e^{-\mu^2/2\omega^2}$ with very large numbers in ${}_1F_1\left(-N-\ell-\tfrac{1}{2}; -\ell+\tfrac{1}{2}; \tfrac{\mu^2}{\omega^2}\right)$. However, what saves the day is the expression (55) which is given in terms of the J-matrix coefficients $T_n$ and $R_n^\pm$ as ratios of $c_n$ and $s_n$. Thus, the small values in $e^{-\mu^2/2\omega^2}$ factor out and cancel whereas the large values in ${}_1F_1\left(-N-\ell-\tfrac{1}{2}; -\ell+\tfrac{1}{2}; \tfrac{\mu^2}{\omega^2}\right)$ will be divided out.

The tridiagonal requirement on the time-independent spinor basis for the inner function space ($n = 0,1,..,N-1$) gives $\xi_n^+(r) = \varphi_n^+(r)$ and $\xi_n^- = \tfrac{\varepsilon+1}{\rho}\varphi_n^-$. Thus, we obtain

$$\zeta_n(r) = a_n x^\ell e^{-x^2/2} \begin{pmatrix} x L_n^{\ell+\frac{1}{2}}(x^2) \\ \tfrac{\lambdabar\omega}{\rho}\left\{ (\kappa+\ell+1) L_n^{\ell+\frac{1}{2}}(x^2) - x^2\left[ L_n^{\ell+\frac{3}{2}}(x^2) + L_{n-1}^{\ell+\frac{3}{2}}(x^2) \right] \right\} \end{pmatrix} \tag{A11}$$

where $\rho$ is a real basis parameter. An alternative, but equivalent, expression that could be more useful for performing integrals using Gauss quadrature approximation is as follows

$$\zeta_n(r) = a_n x^{\ell-1} e^{-x^2/2} \begin{pmatrix} \left(n+\ell+\tfrac{1}{2}\right) L_n^{\ell-1/2}(x^2) - (n+1) L_{n+1}^{\ell-1/2}(x^2) \\ \tfrac{\hbar\omega}{\rho} x \left[ \left(n+\ell+\tfrac{1}{2}\right) L_n^{\ell-1/2}(x^2) + (n+1) L_{n+1}^{\ell-1/2}(x^2) \right] \end{pmatrix}, \quad \kappa > 0 \quad \text{(A12a)}$$

$$\zeta_n(r) = a_n x^{\ell+1} e^{-x^2/2} \begin{pmatrix} L_n^{\ell+3/2}(x^2) - L_{n-1}^{\ell+3/2}(x^2) \\ -\tfrac{\hbar\omega}{\rho} x \left[ L_n^{\ell+3/2}(x^2) + L_{n-1}^{\ell+3/2}(x^2) \right] \end{pmatrix}, \quad \kappa < 0 \quad \text{(A12b)}$$

The overlap matrix of this oscillator basis becomes

$$\langle \zeta_n | \zeta_m \rangle = \delta_{n,m} + \left(\tfrac{\hbar\omega}{\rho}\right)^2 \times$$
$$\left[ \left(2n+\ell+\tfrac{3}{2}\right) \delta_{n,m} + \sqrt{n\left(n+\ell+\tfrac{1}{2}\right)} \delta_{n,m+1} + \sqrt{(n+1)\left(n+\ell+\tfrac{3}{2}\right)} \delta_{n,m-1} \right] \quad \text{(A13)}$$

On the other hand, the matrix representation of the reference Hamiltonian in this basis is

$$(H_0)_{n,m} = \langle \zeta_n | H_0 | \zeta_m \rangle = \delta_{n,m} + (2\rho-1)\left(\tfrac{\hbar\omega}{\rho}\right)^2 \times$$
$$\left[ \left(2n+\ell+\tfrac{3}{2}\right) \delta_{n,m} + \sqrt{n\left(n+\ell+\tfrac{1}{2}\right)} \delta_{n,m+1} + \sqrt{(n+1)\left(n+\ell+\tfrac{3}{2}\right)} \delta_{n,m-1} \right] \quad \text{(A14)}$$

Taking the nonrelativistic limit $H_0 \to 1 + \hbar^2 \hat{H}_0$, where $\hat{H}_0$ is the nonrelativistic reference Hamiltonian (in the oscillator basis as shown in [7] but for a different normalization) gives $\rho = 2$. The finite $N \times N$ matrix representation of the radial potentials and effective mass in the oscillator basis is approximated using Gauss quadrature as

$$\langle \xi_n^+ | V_+ | \xi_m^+ \rangle = (V_+)_{nm}^{\ell+1/2} \quad \text{(A15a)}$$

$$\langle \xi_n^+ | W | \xi_m^- \rangle + \langle \xi_n^- | W | \xi_m^+ \rangle =$$
$$\pm 2 \tfrac{\hbar\omega}{\rho} \left[ \sqrt{\left(n+|\kappa|+\tfrac{1}{2}\right)\left(m+|\kappa|+\tfrac{1}{2}\right)} (U_-)_{n,m}^{|\kappa|\mp\tfrac{1}{2}} - \sqrt{n_\pm m_\pm} (U_-)_{n\pm1,m\pm1}^{|\kappa|\mp\tfrac{1}{2}} \right] \quad \text{(A15b)}$$

$$\langle \xi_n^- | V_- | \xi_m^- \rangle = \left(\tfrac{\hbar\omega}{\rho}\right)^2 \left[ \sqrt{\left(n+|\kappa|+\tfrac{1}{2}\right)\left(m+|\kappa|+\tfrac{1}{2}\right)} (V_-)_{n,m}^{|\kappa|\mp\tfrac{1}{2}} + \sqrt{n_\pm m_\pm} (V_-)_{n\pm1,m\pm1}^{|\kappa|\mp\tfrac{1}{2}} \right.$$
$$\left. + \sqrt{m_\pm \left(n+|\kappa|+\tfrac{1}{2}\right)} (V_-)_{n,m\pm1}^{|\kappa|\mp\tfrac{1}{2}} + \sqrt{n_\pm \left(m+|\kappa|+\tfrac{1}{2}\right)} (V_-)_{m,n\pm1}^{|\kappa|\mp\tfrac{1}{2}} \right] \quad \text{(A15c)}$$

where $n_+ = n+1$ and $n_- = n$ for $\kappa > 0$ and $\kappa < 0$, respectively. The radial function $U_-$ in Eq. (A15b) is defined as $U_-(r) = \tfrac{1}{\omega r} W(r)$. However, one should note that in the oscillator basis the approximation series for the integral in Eq. (50) should be replaced by

$$F_{nm}^\nu \cong \sum_{k=0}^{K-1} \Lambda_{nk}^\nu \Lambda_{mk}^\nu F\left(\sqrt{\eta_k^\nu}/\omega\right) \quad \text{(A16)}$$

**FIGURE CAPTIONS:**

**Fig. 1** : A plot of $\left|1-S^{(N)}(\varepsilon)\right|$ for S-wave scattering as a function of the energy variable $E(\varepsilon)$ and for constant mass ($M = m_0$). The potential matrix is defined by Eq. (59) and Eq. (60) with the physical parameters: $V_0 = 3.0$ a.u., $r_0 = 3.0$ a.u., and with a basis size $N = 50$ and $\omega = 10$ a.u. The solid curve is the relativistic result ($\lambdabar = 0.2$ a.u.) whereas the dashed curve corresponds to the nonrelativistic limit ($\lambdabar = 0.001$ a.u.).

**Fig. 2**: Pure nonrelativistic calculation of the scattering matrix for the same problem as in Fig. 1 (with constant mass) which was performed using the standard nonrelativistic J-matrix method in the Laguerre basis with the same parameters. The agreement with the dashed curve in Fig. 1 is obvious.

**Fig. 3** : The integrated phase shift $\delta(\varepsilon)$ as a function of energy for the same problem as in Fig. 1 (using $\delta = \frac{1}{2} \arg S^{(N)}$). The relativistic (nonrelativistic) result is the solid (dashed) curve.

**Fig. 4** : A reproduction of Fig. 1 but for spatially dependent mass where we took $S_0 = 1.0$ (a.u.). The effect of the spatial variation of mass is very obvious as pointed out at the end of Sec. V. The relativistic (nonrelativistic) result is the solid (dashed) curve.

**Fig. 5** : A reproduction of Fig. 3 for spatially dependent mass with $S_0 = 1.0$ (a.u.). The relativistic (nonrelativistic) result is the solid (dashed) curve.

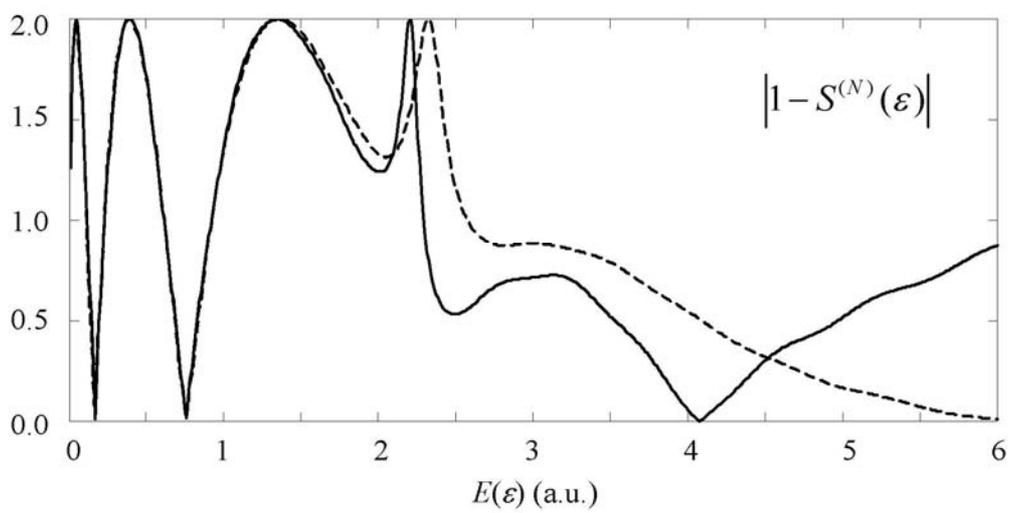

**Fig.1**

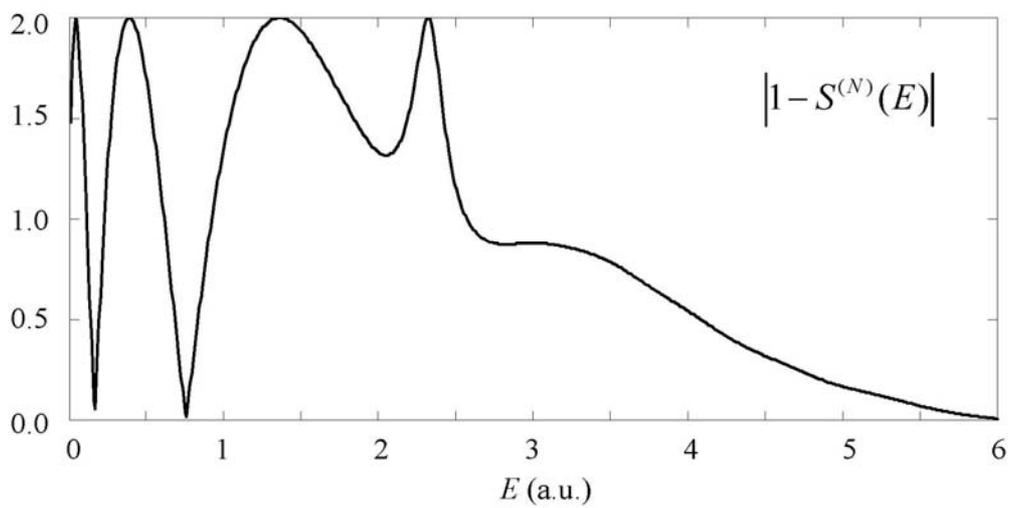

**Fig. 2**

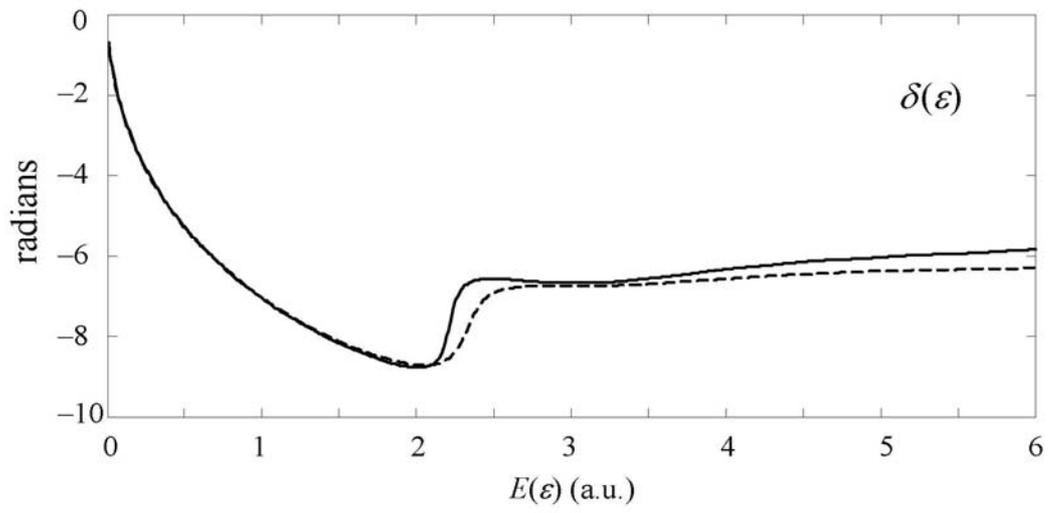

**Fig. 3**

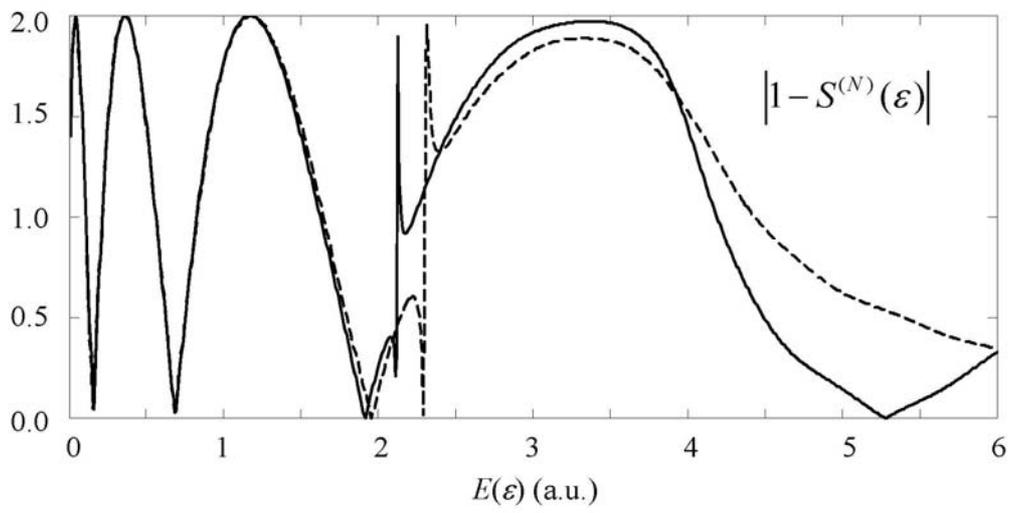

**Fig. 4**

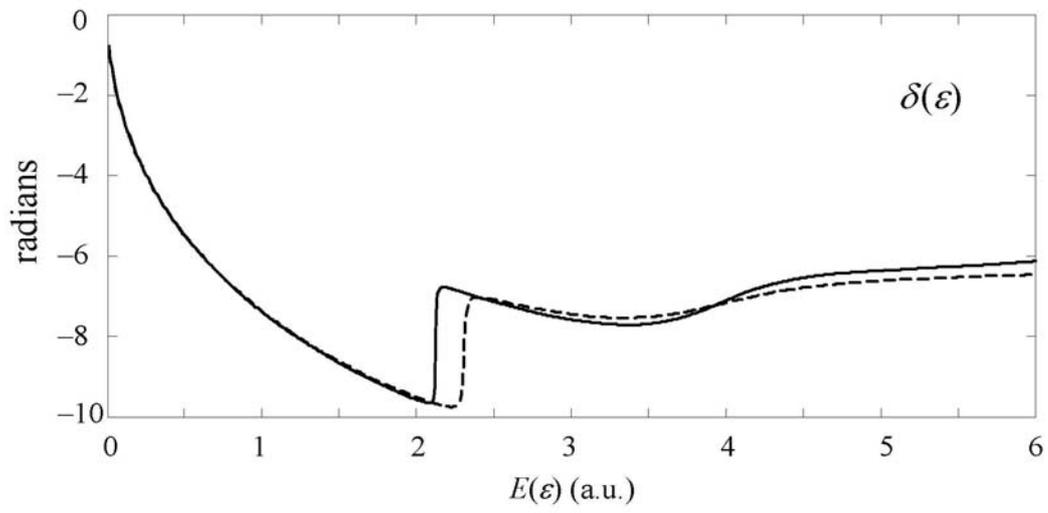

**Fig. 5**